# Room temperature coexistence of large electric polarization and magnetic order in $BiFeO_3$ single crystals.

Delphine Lebeugle[1], Dorothée Colson[1], Anne Forget[1], Michel Viret[1], Pierre Bonville[1], Jean-Francis Marucco[1, 2], Stéphane Fusil[3]

[1]Service de Physique de l'Etat Condensé, DSM/DRECAM, CEA Saclay, 91191 Gif-Sur-Yvette Cedex, France

[2]Laboratoire d'Etude des Matériaux Hors-Equilibre, Université Paris-Sud, Bât. 415, 91405 Orsay Cedex, France

[3]Unité mixte de Physique CNRS-Thalès URA 2464, Route Départementale 128, 91767 Palaiseau, France

**Abstract.** From an experimental point of view, room temperature ferroelectricity in $BiFeO_3$ is raising many questions. Electric measurements made a long time ago on solid-solutions of $BiFeO_3$ with $Pb(Ti,Zr)O_3$ indicate that a spontaneous electric polarization exists in $BiFeO_3$ below the Curie temperature $T_C$=1143K. Yet in most reported works, the synthesised samples are too conductive at room temperature to get a clear polarization loop in the bulk without any effects of extrinsic physical or chemical parameters. Surprisingly, up to now there has been no report of a P(E) (polarization versus electric field) loop at room temperature on single crystals of $BiFeO_3$. We describe here our procedure to synthesize ceramics and to grow good quality sizeable single crystals by a flux method. We demonstrate that $BiFeO_3$ is indeed ferroelectric at room-temperature through evidence by Piezoresponse Force Microscopy and P(E) loops. The polarization is found to be large, around 60 μC/cm², a value that has only been reached in thin films. Magnetic measurements using a SQUID magnetometer and Mössbauer spectroscopy are also presented. The latter confirms the results of NMR measurements concerning the anisotropy of the hyperfine field attributed to the magnetic cycloidal structure.

**I. INTRODUCTION**

Recently, the perovskite-type oxides, which display ferroelectric and magnetic properties, have been the object of a renewed interest as a significant interplay between the two orders would open potential applications in spintronic devices[1]. It has been shown that $BiFeO_3$ is a good candidate as the space group R3c allows the existence of both antiferromagnetic and ferroelectric orders with very high transition temperatures. $BiFeO_3$ is antiferromagnetic below the Néel temperature $T_N$=643K (neutron powder diffraction[2]) with a long range cycloidal spiral incommensurate with the lattice[3] ; it is also ferroelectric below $T_C$=1143K (dielectric measurements[4] and X-Ray single crystal diffraction[2]).

Numerous studies have been performed on $BiFeO_3$ samples and especially, more recently, on thin films[5, 6, 7, 8]. A significant enhancement of the polarization in thin films has been reported in comparison with the bulk and recently, electrical control of antiferromagnetic domains in $BiFeO_3$ films at room temperature has even been evidenced[5, 6]. However it was also argued that epitaxial strain does not enhance the magnetization and polarization in $BiFeO_3$ and that increased thickness-dependent magnetization is not an intrinsic property of fully oxygenated and coherently strained epitaxial $BiFeO_3$ films which exhibit a high electrical resistivity[9]. Furthermore, the conduction in most $BiFeO_3$ ceramics or films seems to be extrinsic, limited by oxygen vacancies and it is easily rectified by replacing some Fe with Mn[10]. We confirm in this paper that the large spontaneous polarization measured in thin films can also be measured on high resistive single crystals at room temperature by means of polarization loop measurements.

Concerning the coupling between magnetic and electric orders, a quadratic magnetoelectric signal has been observed in the bulk[11]. The main limitation for electric measurements is the leakage current which generally prevents the application of a reasonably high electric field on the sample. Hence, very pure samples are required since secondary phases and grain boundaries are responsible for the leakage currents. We present here our improvements to the synthesis of $BiFeO_3$ ceramic and single crystals. The high quality achieved in the single crystals allowed us to measure a polarization loop at room temperature. The ferroelectric behaviour of $BiFeO_3$ is also evidenced by Piezoelectric Force Microscopy measurements. We also demonstrate the absence of weak ferromagnetism in bulk $BiFeO_3$ single crystals, and we present a brief discussion of the effect of a spiral magnetic structure on the $^{57}Fe$ Mössbauer absorption spectra, and a comparison with NMR measurements[12].

## II. SAMPLE PREPARATION

The synthesis of pure polycrystalline $BiFeO_3$ samples is quite subtle because it is necessary to take both kinetic and thermodynamic properties into account as they tend to counteract each other. As can be seen in the phase diagram[13], two impurities can be formed along with $BiFeO_3$. Indeed the high volatility of $Bi_2O_3$ leads to the formation of a Bi-poor phase $Bi_2Fe_4O_9$, but a small excess of $Bi_2O_3$ in the reactants, necessary to compensate for the loss of $Bi_2O_3$, leads to the formation of a Bi-rich phase $Bi_{25}FeO_{39}$. Differential thermal analysis and kinetic investigations[14, 15] show that the reaction in equimolar mixture of $BiFeO_3$ is complex. Below 675°C the reaction is incomplete. From 675°C to 870°C, $BiFeO_3$ decomposes very slowly into $Bi_2Fe_4O_9$, whereas above 870°C $BiFeO_3$ separates rapidly into $Bi_2Fe_4O_9$ and a liquid phase in an incongruent fusion. A previous reported synthesis of $BiFeO_3$ using solid state reactions[16] needs a very large excess of $Bi_2O_3$ (100%) to prevent the formation of $Bi_2Fe_4O_9$, but the disadvantage of this technique is to form a large quantity of $Bi_{25}FeO_{39}$ that cannot be separated from $BiFeO_3$ with a good yield.

We propose here to use the solid state reaction from a stoichiometric mixture by adjusting temperature and time in order to get a complete reaction while preventing the formation of both impurities. This way of preparation differs from the rapid sintering technique previously developed[17, 18] at higher temperatures which can lead to a sample contamination by the crucible. The details of our preparation are as follows: a stoichiometric $Bi_2O_3$:$Fe_2O_3$ mixture is ground finely and sintered in air for 15h at 800°C using an alumina crucible. It should be noted that the crucible can only be used for one or two experiments; after this the porosity of the crucible may cause a loss of Bi. Powder X-ray diffraction data were collected at room temperature using a "D8 Advance" diffractometer (Bruker-axs) with $Cu_{K\alpha}$ radiation (λ = 1.5418 Å) and a scan rate of 0.02° per 20s. Intensities of 18 reflections in the range 10°<2θ<70° were collected with an energy dispersive detector ("Sol-X" Bruker). The analysis shows no traces of $Bi_2Fe_4O_9$ but a small amount of $Bi_{25}FeO_{39}$ (figure 1). This parasitic phase can be leached in 10% diluted $HNO_3$. With this technique, we obtain a dark-brown single-phase polycrystalline $BiFeO_3$ which has been used to grow epitaxial thin films[19].

The possible non-stoichiometry of $BiFeO_{3+\delta}$ was studied by thermogravimetry in a controlled atmosphere between 25 and 800°C. The experimental set up has been previously described[20]. It consists of a Setaram B60 thermobalance with a working sensibility of 0.015 mg corresponding to $\Delta\delta = 5.10^{-4}$ for a mass sample of 1.5 g. A Chevenard furnace was used to adjust the temperature. The $BiFeO_3$ powder was directly heated in a recrystallized alumina crucible under pure oxygen and then argon atmospheres. The oxygen pressure in argon corresponds to $10^{-5}$ atm. No composition variation was detected in the studied temperature range both under oxygen and argon, characterizing a stoichiometric compound. The use of more reducing atmospheres ($CO$-$CO_2$ or $H_2$-$H_2O$), corresponding thermodynamically to $Fe^{2+}$ or Bi metal, would lead to the decomposition of the perovskite phase.

In order to study the intrinsic properties of $BiFeO_3$, we have also developed a procedure to grow single crystals of a millimetric size needed for physical measurements. More specifically, electrical measurements require producing platelets with one reduced dimension and two large ones, hence maximising the electric field which can be applied on the material as well as optimising the charges on both sides (which are proportional to the sample area). Furthermore, grain boundaries being more conducting than $BiFeO_3$ crystallites, it is important to minimise their density in order to be able to apply a large enough electric field on the samples. The single crystals were grown in air from a $Bi_2O_3$-$Fe_2O_3$ flux with a mole ratio 3.5:1 using an alumina crucible. A fine and homogeneous mixture was obtained from approximately 75 g quantity of the highly pure oxides $Bi_2O_3$ and $Fe_2O_3$ stirred with deionised water using a grinding mill (Attritor). In order to increase the surface-to-volume ratio, we put a thin layer of the melt (2cm) in an alumina crucible 6 cm in diameter. The mixture was heated up to 850°C at a rate of 150°C/h, held for 4 h at this temperature and was cooled down slowly to 750°C; at this point the furnace was turned off. The crystals grown in this way are millimetre sized black platelets (figure 2.a). They can be extracted from the solid flux mechanically. Electron microprobe analysis confirms the stoichiometry of $BiFeO_3$.

The ferroelectric domains have been observed under a polarizing microscope in reflectivity mode (figure 2.b). The contrast between bright domains (which correspond to a direction of the spontaneous polarization) and dark domains (which corresponds to the opposite one) comes from the birefringence property of the ferroelectric crystal. Then we are able to select from the batch crystals of single ferroelectric domain type (figure 2.c)

## III. SINGLE CRISTAL X-RAY ANALYSIS

Four-circle X-ray diffraction data were collected at room temperature on a single crystal by using a Kappa X8 APPEX II Bruker diffractometer with graphite-monochromated $Mo_{K\alpha}$ radiation ($\lambda$ = 0.71073 Å). Lattice parameters were determined from 1230 reflections in the range of 2.64°<θ<32.2° using a two-dimensional detector. The data were corrected for Lorentz polarization, and absorption effects. The structure was solved by direct methods using SHELXS-97[21] and refined against $F^2$ by full-matrix least-squares techniques using SHELXL-97[22] with anisotropic displacement parameters for all atoms. All calculations were performed by using the Crystal Structure Crystallographic software package WINGX[23]. The crystals have a rhomboedrally distorted perovskite-type cell with lattice constants $a_{hex}$=5.571Å and $c_{hex}$=13.868 Å at room temperature. The space group is determined as R3c with six formula units in the unit cell (figure 3). Data collection and refinement parameters are given in Table 1. They are close to those obtained by F. Kubel and H. Schmid by X-Ray diffraction ($a_{hex}$= 5.579Å, $c_{hex}$= 13.869Å) [24]. This set up allows us to orient the crystal correctly for magnetic and electric measurements. The largest side of the crystal corresponds to the $(012)_{hex}$ plane (the $(110)_{cub}$ plane). The axis perpendicular to this plane makes an angle of 54°44' with the c-axis. This crystal orientation renders the physical measurements quite difficult, because the polarization axis $[001]_{hex}$ (i.e $[111]_{cub}$) and the interesting magnetic diffraction plane $(1\text{-}10)_{hex}$ (i.e. $(10\text{-}1)_{cub}$) which contains the satellite peaks due to the cycloidal structure, are not easy directions of the shape of the crystal.

## IV. FERROLECTRIC PROPERTIES

The ferroelectric character of polycrystalline samples at room temperature has already been evidenced by the observation of saturation polarization loops[17, 18, 25]. However, contributions from leakage currents and the samples own capacitance makes it difficult to quantify the intrinsic polarization of the ceramics. As a result, very low values of the polarization have generally been inferred from the P(E) loops (about 6 $\mu m.cm^{-2}$). Indeed the conductivity of $BiFeO_3$ is not negligible at room temperature, presumably due to impurities at the grain boundaries that lead to high porosity. It should be noted that another way to obtain a higher dielectric constant and a lower dielectric loss factor is to make solid-solutions of $BiFeO_3$ with another $ABO_3$ perovskite like $BiFeO_3$-Pb(Ti,Zr)$O_3$[4]. The measurements are made possible that way, but a proper quantification of the polarization and coercive fields is also difficult. It is easier to measure epitaxial thin film and much larger values (about 60 $\mu m.cm^{-2}$) have been reported[4]. The general belief is that this tenfold enhancement of the remanent polarization in thin films is due to a structural modification (to a tetragonal crystal structure) coming from compressive stress imposed by the substrate. Here, we show that the large values found in films are intrinsic to $BiFeO_3$ phase as demonstrated by the room-temperature polarization we measure in high purity single crystals. These large values of polarization are actually consistent with theoretical predictions[26]. From Piezoelectric Force Microscopy measurements, the remnant out-of-plane piezoelectric coefficient $d_{33}$ in these films is about 70 pm/V.

In this article, we report on our work on the pure bulk compounds. Our efforts to compact sintered samples by isostatic pressure (2500 bars) lead to an increase in resistivity, but this was still not sufficient to obtain a clear polarization loop at room temperature. To overcome these persistent problems, we have chosen to synthesize single crystals. Surprisingly, up to now only very few measurements have been reported on single crystals, including a minor hysteresis loop at 77K[27]. The measured spontaneous polarization was 3.5 $\mu C.cm^{-2}$ along the $[101]_{hex}$ direction, which would represent 6.1 $\mu C.cm^{-2}$ in the $[001]_{hex}$ direction. According to the authors, the difficulty to obtain a well saturated P(E) loop comes from the high conductivity of $BiFeO_3$ for temperatures above 190K ($\rho(300K) \approx 0.2\text{-}1.10^5$ $\Omega m$ at 1kHz[28]). This is probably due to impurities on the surface or inclusions in the crystals. Clearly, there is a need for more measurements on pure crystals.

*IV.A Piezoelectric Force Microscopy*

The ferroelectric behaviour of a 40 μm thick $BiFeO_3$ single crystal obtained by the previously described method was assessed at room temperature by Piezoelectric Force Microscopy (PFM). The crystal was polished to a thin platelet to ensure that the bias range available in the PFM set up was sufficient to reverse the polarization. The $BiFeO_3$ platelet was fixed with silver paint acting as a bottom electrode while a bias is applied on the conductive atomic force microscope tip used as the top electrode. Imaging of the domains is carried out by applying a small AC bias on the bottom electrode (~10 times lower than the coercive voltage) giving rise to the converse piezoelectric effect detected trough lock-in techniques. The Out of Plane phase component (OP-PFM) of the piezoresponse is homogeneous over areas as large as 20x20$\mu m^2$ demonstrating that the sample is fully saturated (figure 4.a). The In Plane phase component (IP-PFM) exhibits two levels of contrast shifted by 180° one from the other most likely because of sample roughness at a submicrometric scale (figure 4.b).

Figure 5.a shows the topography of the thinnest region of the single crystal (6x6 $\mu m^2$ area). By applying a DC bias to the scanning tip, it is possible to draw ferroelectric patterns: "up" and "down" stripes (0.5x4$\mu m^2$) were written with a positive and negative DC bias respectively. The pattern written can be subsequently imaged as shown in figure 5.b revealing the written stripes.

A local hysteretic cycle can be measured through the local piezoelectric response versus applied voltage. The typical cycle shown in figure 6 unambiguously confirms the ferroelectric nature of the single crystal at room temperature. A quantitative estimate of the piezoresponse and of the coercive field by this technique requires the elaboration of plane capacitors. In the experiments reported here, the effective electric field is amplified by the tip effect and hence not precisely known. Taking a rough estimate for the tip radius at its apex (around 100nm), we can infer a coercive field of a few tens of kV/cm.

*IV.B Charge current versus applied voltage*

Larger crystals have been used for polarization loop measurements in a capacitor geometry by a standard 'macroscopic' method which consists in measuring the current flowing in a simple resistive circuit as a function of the voltage applied to the sample (I(V) characteristic). Any change of the sample polarization induces a transfer of charges which is measured by a pico-ampere meter. This current is the time derivative of the charge, which is directly related to the polarization cycle. Experimentally, one measures a small leakage current (due to the fact that the sample has a finite resistance) superimposed to the derivative of the ferroelectric hysteresis cycle. For the measurements presented here, electrodes were apposed using silver paste applied on the two major faces of the sample (which correspond to the plane $(012)_{hex}$) and copper wires were used to apply the voltage across the crystal. The electric field dependence of the charge current (I-E) was measured using a Keithley picoamperemeter and voltage source. The maximum electric field that could be applied on our samples is 125kV/cm, which proved high enough to observe a hysteresis cycle. As shown in figure 7.a, the raw data are composed of a background leakage current superimposed to the relevant signal composed of two peaks, either positive or negative, which are due to the charges flowing through the circuit when the sample polarization reverses. The resistance of the sample is very high ($\rho$(300K,100V) ~ $6.10^{10}$ Ω.cm) hence allowing us to observe a large polarization loop. In order to precisely extract the charge current from the measured leakage current, we fitted the latter to a $V^3$-dependence typical from leakage current. When integrating the signal, a hysteresis cycle can be reconstructed as shown in figure 7.b. The relevant quantities of spontaneous polarization and coercive fields, in the direction of the applied electric field, can be easily and reliably extracted. We found for our $BiFeO_3$ single crystal at room temperature a remnant polarization $P_{(012)}$ of 35 μC.cm$^{-2}$ and a coercivity of 15 kV/cm. Taking into account that our measurement is not in the exact easy direction (but canted by 54°44'), the inferred full saturation polarization along the [001] direction is close to 60 μC.cm$^{-2}$. This value is in good agreement with theoretical predictions[26] and thin film data[5], but it contradicts previous room-temperature measurements in bulk systems giving polarizations an order of magnitude lower. Our results clearly demonstrate that the intrinsic polarization of the $BiFeO_3$ crystals is indeed that expected from calculations.

## V. MAGNETIC PROPERTIES

$BiFeO_3$ presents a G-type antiferromagnetic structure ($T_N$=643K)[2] with a long range cycloidal spiral[3] (figure 8). The propagation vector **q** is along the [110] direction and lies in the plane of spin rotation (1-10). It is an incommensurate magnetic order with a period length of 62 nm.

We measured magnetization as a function of applied magnetic field and temperature using a Superconducting QUantum Interference Device (SQUID Quantum Device). The magnetization curve obtained at 300K on the powder sample (figure 9) is linear with the field, which is typical for an antiferromagnetic arrangement of the $Fe^{3+}$ magnetic moments, with a slope $\chi = 3.59\ 10^{-3}$ emu/mol. In the molecular field approximation, the 0K perpendicular AF susceptibility is given by: $\chi_\perp = M_0/H_E$, where $M_0$ is the magnetization of one sublattice and $H_E$ is the first neighbour exchange field[29]. The latter is linked to the Néel temperature by: $k_B T_N = (2/3N).M_0.H_E$ and the powder susceptibility is: $\chi_{AF} = (2/3)\ \chi_\perp$. Using $T_N = 643$ K, we obtain $\chi_{AF} = 3.40\ 10^{-3}$ emu/mol, which is close to the measured slope. Considering this result, the AF susceptibility does not seem to be affected by the cycloidal structure, probably because the period (620Å) is very long in comparison with the lattice parameters (about 5Å).

The isothermal magnetization curves M(H) of an unleached single crystal (sample S1) of $BiFeO_3$ at several temperatures, with magnetic field applied perpendicularly to the (012) plane, are shown in figure 10. At 2K and 5K, we observe an extra contribution, with downwards curvature, superposed on the AF linear field dependence. The magnetization curves versus temperature M(T) under a field of 1 T are shown in figure 11, for the unleached crystal S1 (blue points) and for a crystal leached in diluted $HNO_3$ (sample S2, red points). One observes a constant magnetization at high temperature, and an upturn below 30K for sample S1, and below 10K for sample S2. For sample S1, the non-linear M(H) observed at 2 and 5K, and the low temperature upturn of M(T) can be consistently interpreted in terms of respectively a Brillouin law and a Curie law due to 1% mol. paramagnetic $Fe^{3+}$ ions. It is very likely that these $Fe^{3+}$ ions belong to the $Bi_{25}FeO_{39}$ phase formed from the excess of $Bi_2O_3$ used for the crystal growth. In the unleached sample, this impurity phase amounts thus to 16% of the $BiFeO_3$ mass. The same analysis made on the M(T) curve of sample S2 shows a strongly reduced impurity content (6% wt.). This shows that $HNO_3$ leaching is efficient in order to remove the $Bi_{25}FeO_{39}$ impurity and that the weak ferromagnetism reported in reference 25 in polycrystalline $BiFeO_3$ is in fact due to the presence of a small amount of $Bi_{25}FeO_{39}$ impurity. Hence, our measurements confirm that in pure $BiFeO_3$ weak ferromagnetism vanishes because of the antiferromagnetic cycloidal structure[30].

## $^{57}$Fe MÖSSBAUER SPECTROSCOPY

The first complete study of $BiFeO_3$ using Mössbauer spectroscopy on $^{57}$Fe was done by C. Blaauw and F. van der Woude[31]. In all the spectra well below $T_N$, they observe a differential (or inhomogeneous) broadening of the lines, which they tentativley assign to the presence of different non-equivalent Fe sites in the crystal structure. The motivation of our Mössbauer study was to determine whether the observed spectral asymmetry is linked with the spiral magnetic structure, and to examine the relation with the anisotropy observed in the $^{57}$Fe zero field NMR spectrum of $BiFeO_3$[12, 32]. As Bi has a high electronic absorption coefficient for the Mössbauer 14.4 keV γ-rays, an absorber was made with Fe 30% enriched in $^{57}Fe$. The Mössbauer spectrum was recorded using a constant acceleration drive spectrometer and a commercial $^{57}Co$:Rh $\gamma$-ray source. For $^{57}Fe$, 1 mm/s corresponds to 11 MHz.

The spectrum obtained at room temperature, shown in figure 12, is very similar to the spectrum at 80 K shown in reference 31. It consists of a 6-line magnetic hyperfine pattern, with lines somewhat broadened with respect to the minimal experimental linewidth and presenting a sizeable asymmetry, i.e. the spectrum shows inhomogeneous line broadenings. A 6-line spectrum is characteristic of magnetically ordered $Fe^{3+}$ moments, and it is due to the presence of a hyperfine magnetic field $\mathbf{H}_{hf}$ acting on the nuclear spin. For $Fe^{3+}$, $\mathbf{H}_{hf}$ is antiparallel to the magnetic moment $\mathbf{m}$ and its room temperature value in $BiFeO_3$ is close to 50 T. A small electric quadrupolar hyperfine interaction $\mathcal{H}_Q$, due to the coupling of the nuclear quadrupole moment Q of the excited nuclear state (spin I=3/2) with the electric field gradient (EFG) at the nucleus site, is also present (the ground nuclear state has spin 1/2, and thus no quadrupole moment). We will show in the following that a spiral magnetic structure leads to line broadenings, due to the small modulation of the excited state hyperfine energies arising when the hyperfine field rotates with respect to the principal axis OZ of the EFG tensor $V_{ij}$. The total hyperfine Hamiltonian for the excited nuclear state writes:

$$\mathcal{H}_{hf} = - g_n.\mu_n.\mathbf{H}_{hf}\,.\,\mathbf{I} + \mathcal{H}_Q$$

where $g_n$ is the nuclear gyromagnetic factor and $\mu_n$ the nuclear Bohr magneton. In $BiFeO_3$, the point symmetry at the Fe site is trigonal (a threefold axis along the crystal **c** axis), and thus the principal axis OZ is the crystal **c** axis. Due to the spiral magnetic structure, the Fe moments **m** rotate in the hexagonal (1-10) plane containing OZ. For a given orientation $\Theta$ of **m** (or $\mathbf{H}_{hf}$) with respect to OZ, one can calculate the hyperfine energies up to second perturbation order of $\mathcal{H}_Q$ with respect to the magnetic interaction for each of the 6 lines (labelled by the index i). The angle dependent part writes:

$$\delta^2 E_i\,(\Theta) = \varepsilon_i\, 3\Delta E_Q/4\,(1 + \beta_i\, \Delta E_Q/h\, \sin^2\Theta)\cos^2\Theta, \quad i = 1 \text{ to } 6,$$

where the quadrupolar coupling parameter is $\Delta E_Q = eQV_{ZZ}/2$, $V_{ZZ}$ being the principal value of the EFG tensor, $h=1/2\, g_n\, \mu_n\, H_{hf}$, $\varepsilon_i = 1$ for the two outer lines and -1 for the other lines, and $\beta_i$ is a coefficient which is different for each line. As $\Theta$ varies, the variation of $\delta^2 E_i$ gives rise to a spread in hyperfine energy values, i.e. to a line broadening. For the present case of $BiFeO_3$ of a spiral structure with the principal axis OZ lying in the plane of the moments, and of a very small propagation vector, the angle $\Theta$ varies continuously between 0 and $2\pi$; it is then readily seen that the range of values scanned by $\delta^2 E_i$ is independent of $\beta_i$, i.e. it is the same for all the lines, yielding homogeneous broadenings. More generally, it can be shown that a spiral magnetic structure, either commensurate with a small propagation vector or incommensurate, yields homogeneous line broadenings in a $^{57}Fe$ Mössbauer spectrum when OZ lies in the plane of the magnetic moments; if OZ is perpendicular to this plane, no broadenings occur up to 2nd perturbation order. Inhomogeneous broadenings show up for any other orientation of the principal EFG axis with respect to the plane of the moments. Therefore, the spectral asymmetry observed in $BiFeO_3$ cannot be accounted for by the presence of the spiral magnetic structure; however, our simulations show that the spiral structure alone accounts for the average line broadening, with a value of the quadrupole coupling parameter $\Delta E_Q = 0.48$ mm/s, very close to the value obtained just above $T_N$ in reference 31 (0.44 mm/s).

Recently Zalesski *et al.*[12,32] made zero field $^{57}Fe$ NMR measurements in $BiFeO_3$ below $T_N$ and noted a peculiar shape of the hyperfine spectrum, assigned to an anisotropy of the hyperfine resonance frequency when the hyperfine field rotates in the (1-10) plane. As $^{57}Fe$ NMR concerns the ground nuclear state with no quadrupole moment, this anisotropy cannot arise from the mechanism discussed above for the Mössbauer spectra. According to reference 31, this anisotropy concerns the magnetic hyperfine interaction alone and it is not related to an anisotropy of the $Fe^{3+}$ moment magnitude along the spiral. The angle dependent NMR frequency writes:

$$\omega(\Theta) \approx \omega_{//} \cos^2\Theta + \omega_{\perp} \sin^2\Theta ,$$

and the NMR lineshape is well reproduced with an anisotropy parameter $e=\omega_{//} /\omega_{\perp} \approx 0.990$. Following the NMR study, we introduced an anisotropic variation of the hyperfine field magnitude as the moment rotates in the (1-10) plane:

$$H_{hf}(\Theta) \approx H_{//} \cos^2\Theta + H_{\perp} \sin^2\Theta.$$

Then the spectral asymmetry is very well reproduced with an anisotropy parameter $h = H_{//} /H_{\perp} \approx 0.987$, very close to the NMR value 0.990 (solid line in figure 12).

Therefore, the line broadenings and spectral asymmetry of the Mössbauer spectra in the spiral magnetic phase of $BiFeO_3$ are due to different causes: the broadenings arise from the slight modulation of the hyperfine energies as the magnetic moment rotates with respect to the principal axis of the EFG tensor (crystal **c** axis), and the asymmetry stems from an intrinsic anisotropy of the magnetic hyperfine interaction at a site with trigonal symmetry.

## VI. CONCLUSION

In summary, we showed that highly pure single crystals of $BiFeO_3$ can be grown by a flux method. Their high resistivity allows to observe a clear hysteretic polarization loop at room temperature which confirms the ferroelectricity of $BiFeO_3$. We infer a large intrinsic electric polarization (~ 60 $\mu C.cm^{-2}$) like that predicted from the perfect $BiFeO_3$ unit cell. We conclude that large polarizations are intrinsic to this material and do not stem from some particular property of the thin films. Our magnetic measurements reveal that $BiFeO_3$ has a pure antiferromagnetic response, without any trace of weak ferromagnetism in contradiction with previous reports on polycrystalline samples. Finally, we show that the slight anisotropy of the magnetic hyperfine interaction introduced for interpreting NMR spectra in $BiFeO_3$, perfectly explains the $^{57}$Fe Mössbauer spectra. As to magnetoelectric measurements at room temperature, they are yet in progress.

## ACKNOWLEDGMENTS

The authors are grateful to P. Monod for the possibility of using the SQUID Quantum Device magnetometer and for useful discussions. We want to thanks S. Poissonnet for performing electron microprobe chemical analysis and also R. Guillot for X-Ray diffraction measurements on single crystals. D. Colson wants to acknowledge the Conseil Régional de l'Ile de France (Sésame 2002-2006) and the MIIAT project "Matériaux à Propriétés Remarquables" for their financial supports. This research is supported by the Agence Nationale de la Recherche, project "FEMMES" NT05-1_45147.

# Figure 1

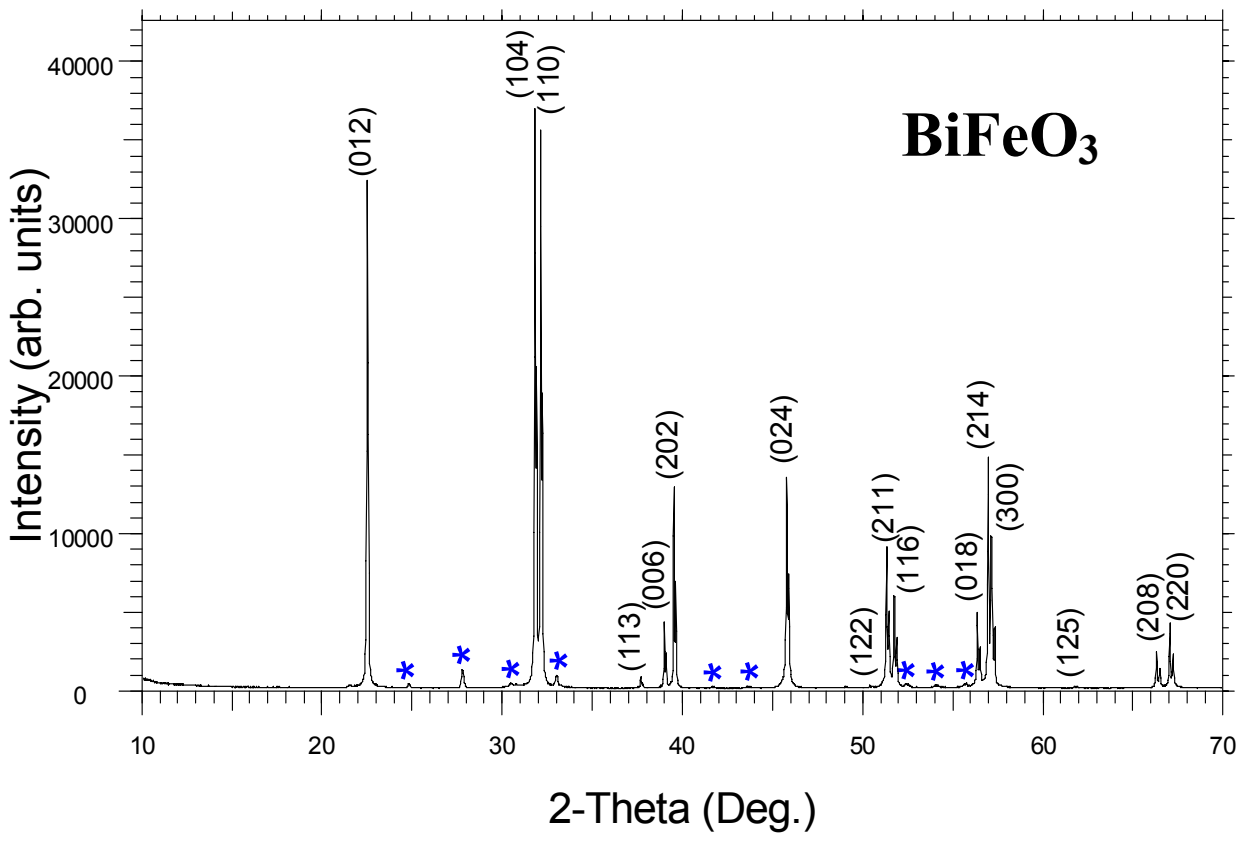

# Figure 2

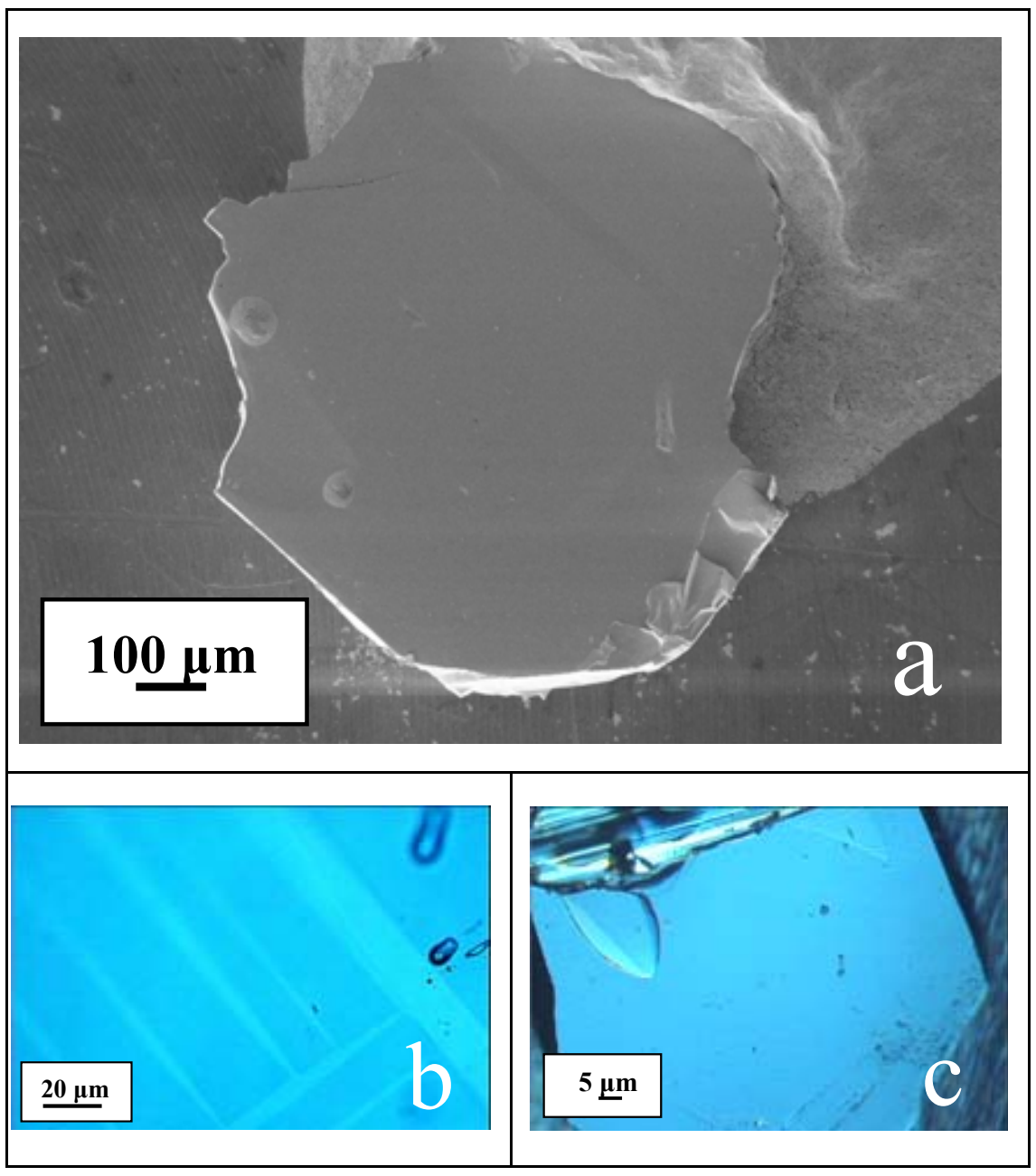

**Figure 3**

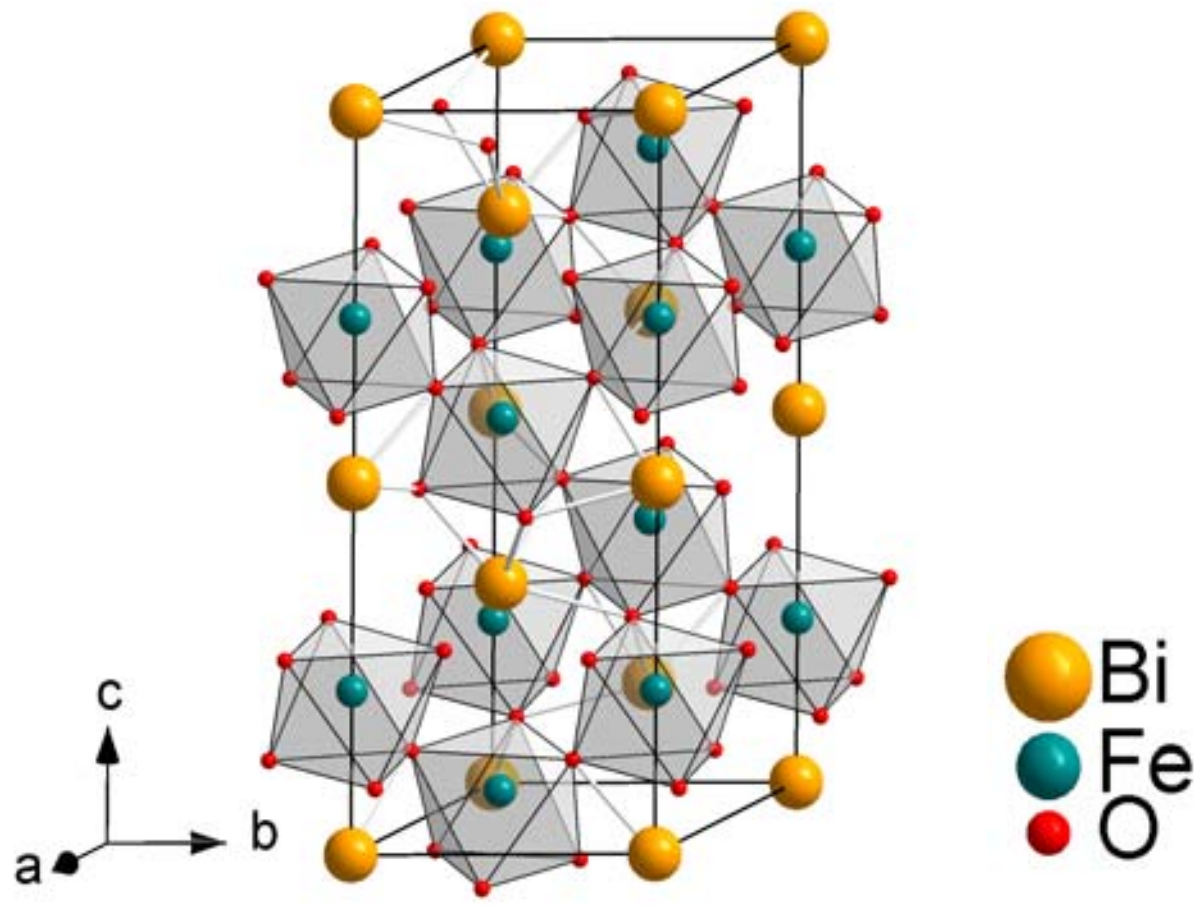

**Figure 4**

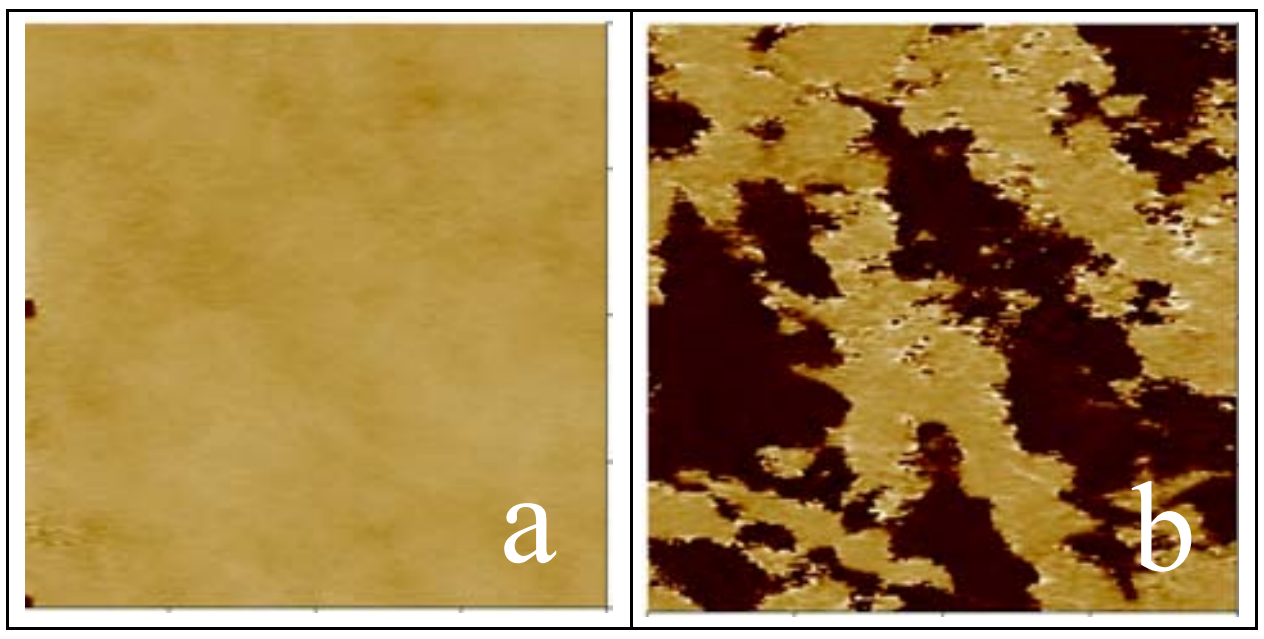

# Figure 5

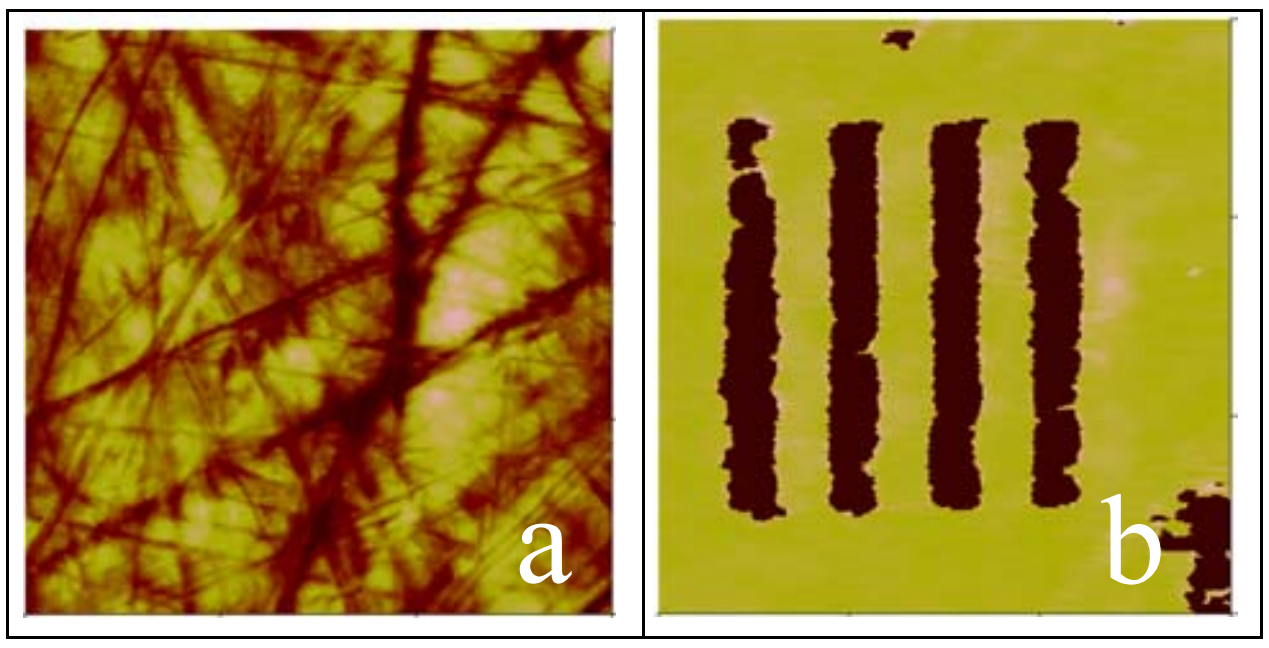

# Figure 6

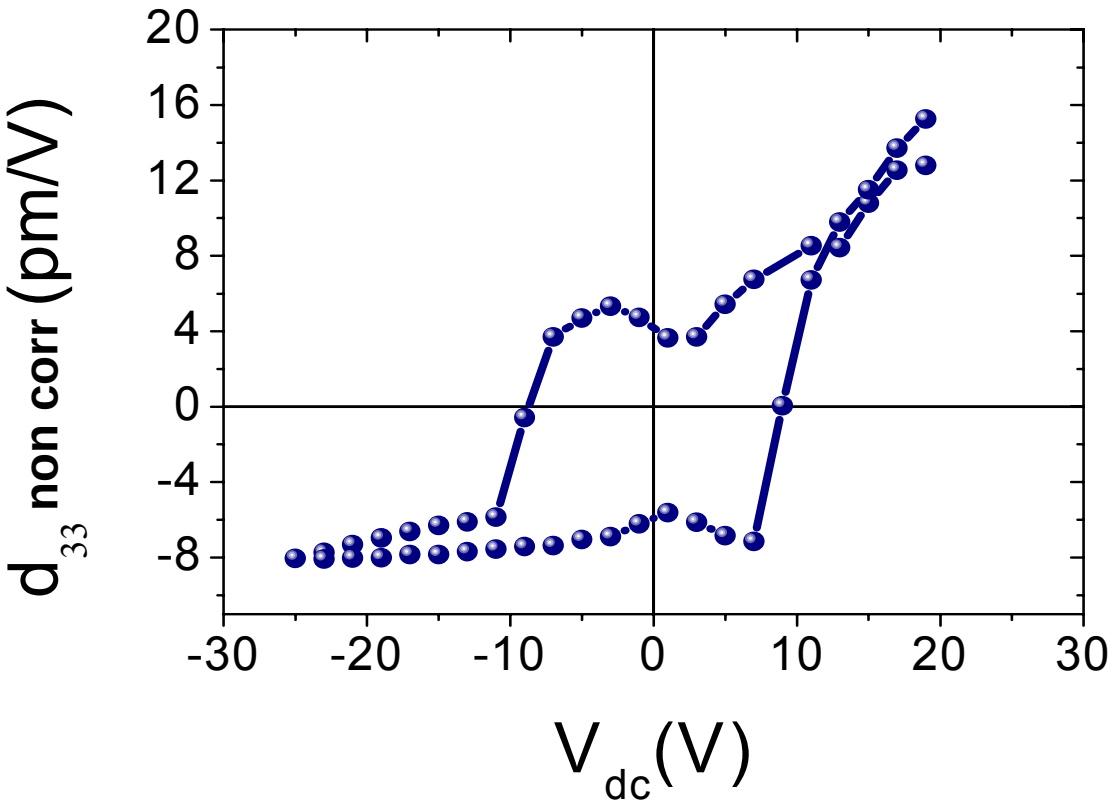

# Figure 7

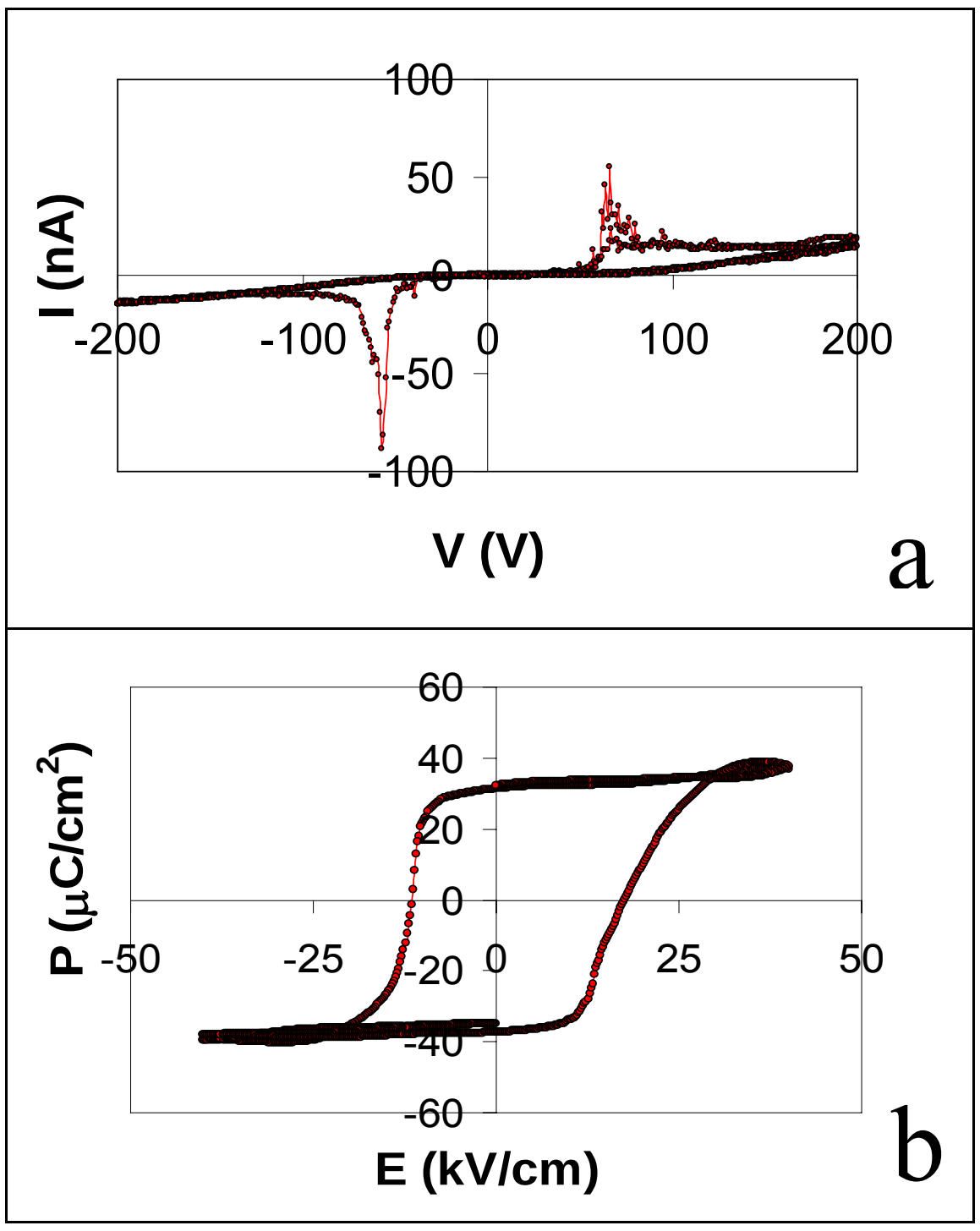

# Figure 8

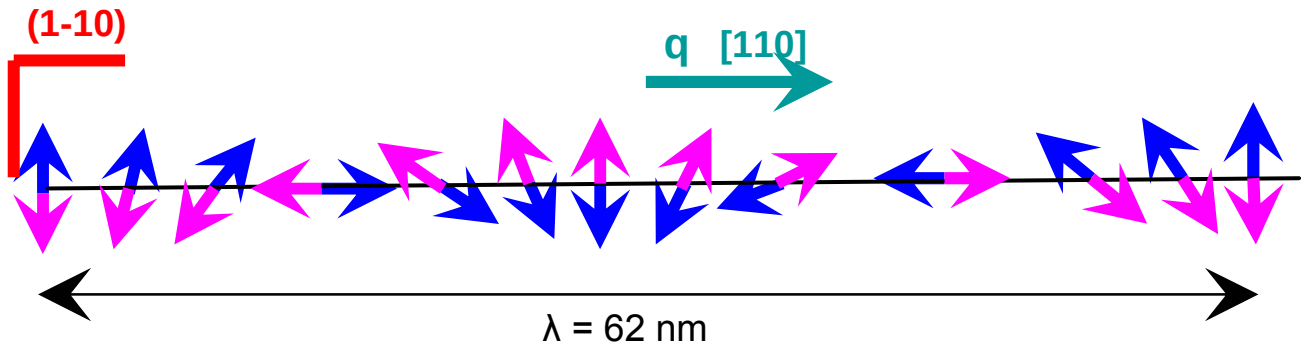

# Figure 9

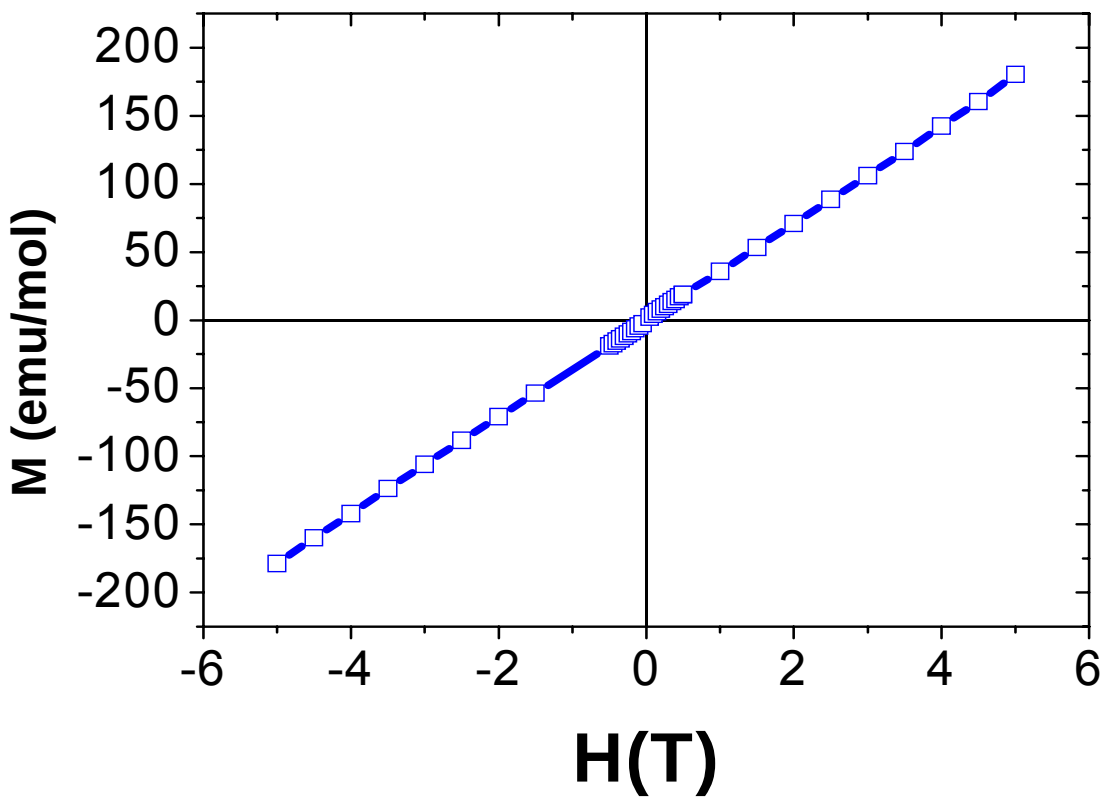

# Figure 10

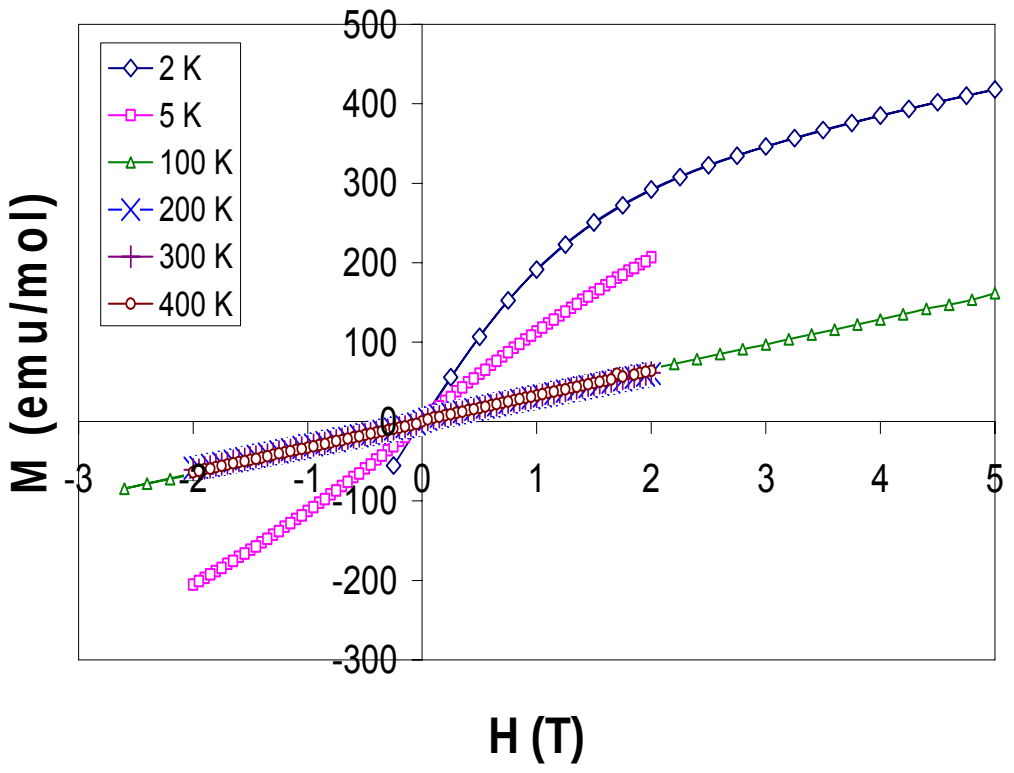

# Figure 11

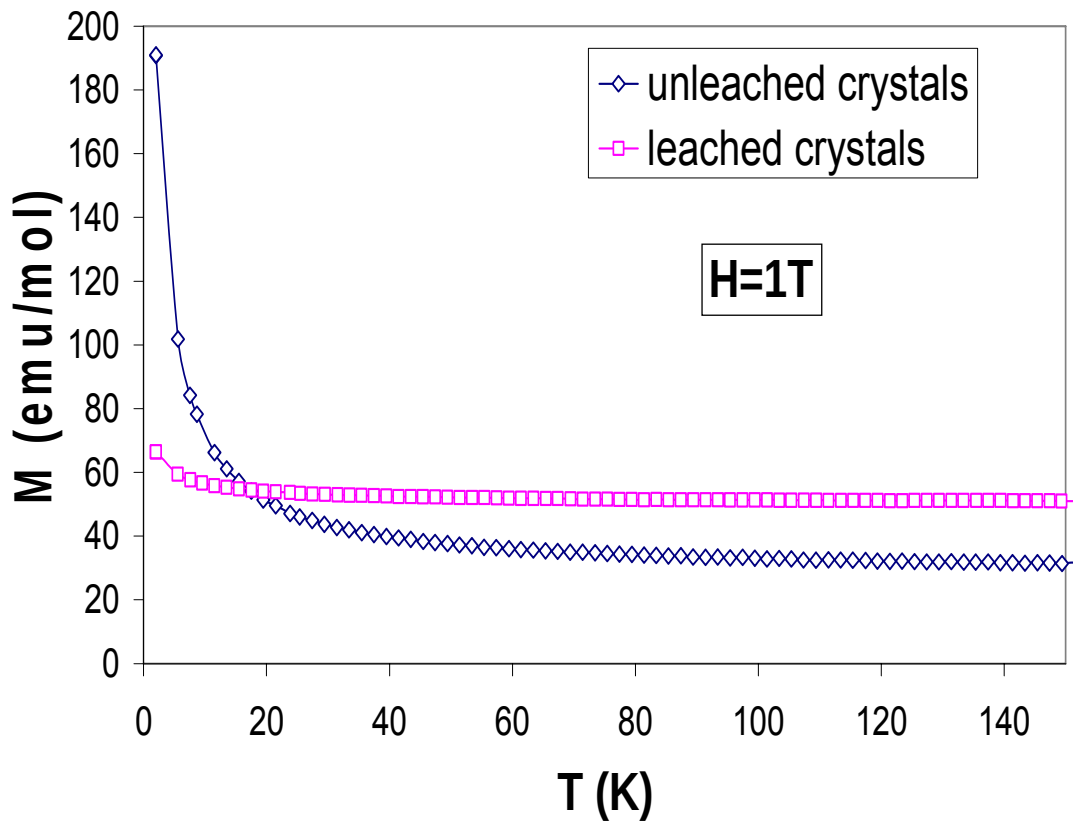

# Figure 12

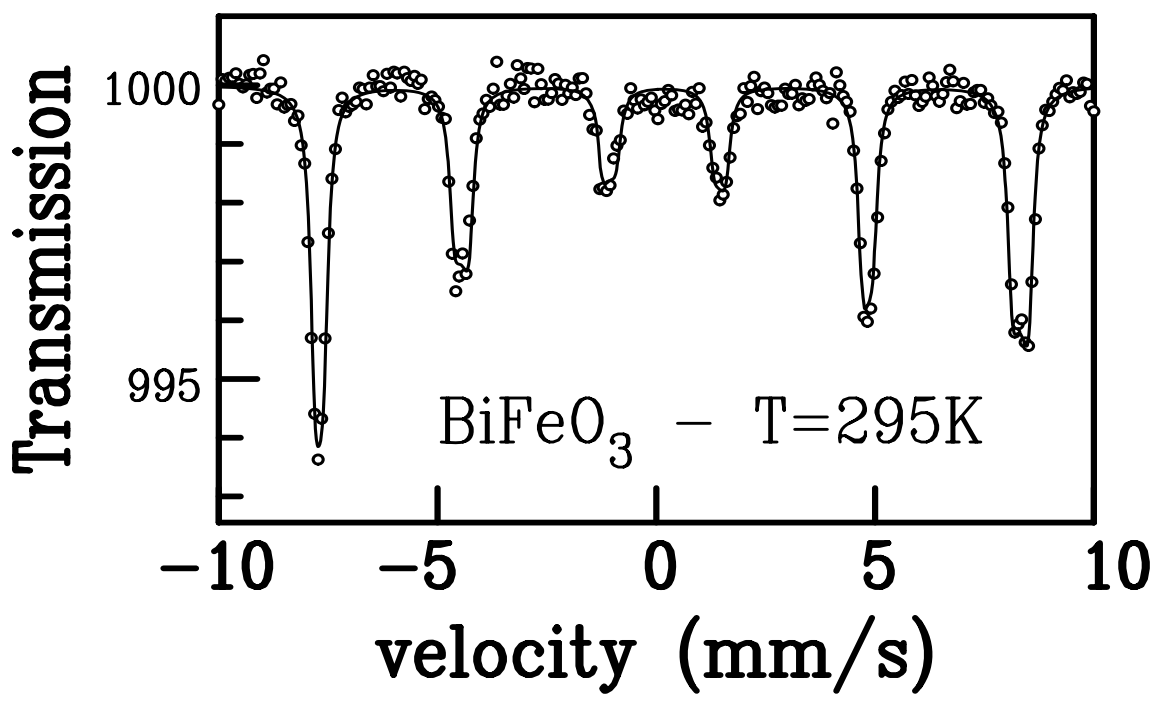

# Table 1

| | |
|---|---|
| Formula | $BiFeO_3$ |
| FW ($g.mol^{-1}$) | 312.825 |
| Crystal system | Rhombohedral |
| Space group | R 3 c |
| $a_{hex}$ (Å) | 5.571(5) |
| $b_{hex}$ (Å) | 5.571(5) |
| $c_{hex}$ (Å) | 13.858(5) |
| α (°) | 90 |
| β (°) | 90 |
| γ (°) | 120 |
| V ($Å^3$) | 372.53(1) |
| Z | 6 |
| ρ ($g.cm^{-3}$) | 8.33 |
| Crystal Size ($mm^3$) | 0.3 x 0.2 x 0.02 |
| F(000) | 798 |
| λ (Kα Mo) (Å) | 0.71073 |
| T (K) | 293(2) |
| μ ($mm^{-1}$) | 76.32 |
| θ range [° min-max] | 2.64 - 52.66 |
| Number of data collected | 12 952 |
| Number of unique data | 414 |
| R(int) | 0.0676 |
| Number of variable parameters | 16 |
| Number of observed reflections[a] | 344 |
| R[b] observed | 0.0763 |
| Rw[c] observed | 0.2232 |
| Goodness of fit S | 1.115 |
| $(\Delta/\sigma)_{max}$ | 0.091 |
| $(\Delta/\rho)_{min}$ (e $Å^{-3}$) | -2.558 |
| $(\Delta/\rho)_{max}$ (e $Å^{-3}$) | 3.651 |

[a] Data with $F_o>4\sigma(F_o)$ [b] $R = \Sigma||F_o|-|F_c||/\Sigma|F_o|$ [c] $Rw = [\Sigma w(|F_o^2|-|F_c^2|)^2/\Sigma w|F_o^2|^2]^{1/2}$.

# Figures captions

**Figure 1** : Powder X-Ray diffraction pattern of $BiFeO_3$ at 300K using $Cu_{K\alpha}$ radiation.The asterisks correspond to the $Bi_{25}FeO_{39}$ impurity. $BiFeO_3$ is present as the major phase.

**Figure 2.a** : SEM image of a $BiFeO_3$ single crystal platelet grown by slow cooling from a $Bi_2O_3/Fe_2O_3$ flux. The crystal size is about 1.4x1.6x0.04 $mm^3$.

**Figure 2.b** : (Color online) A photograph of one part of a $BiFeO_3$ single crystal examined under a polarizing microscope with a polychromatic light. The contrast between bright and dark regions is due to positive and negative orientations of the spontaneous polarization.

**Figure 2.c** : (Color online) A photograph of another $BiFeO_3$ single crystal examined under a polarizing microscope with a polychromatic light. Only one single ferroelectric domain is visible.

**Figure 3** : (Color online) Drawing of the hexagonally distorted perovskite-type $BiFeO_3$ cell using Diamond 3.1 software. Oxygen octahedra are top-connected and oppositely rotated around the threefold axis. All the atomic displacements responsible for the appearance of a dipolar moment occur along the threefold axis.

**Figures 4 :** (Color online) PFM images (20x20 $\mu m^2$ area) of a $BiFeO_3$ polished single crystal.

**Figure 4.a:** Out-of-plane PFM phase homogeneous image.

**Figure 4.b:** In-plane PFM phase image: the two tones are shifted by 180° one from each other as expected for opposite in plane components.

**Figures 5 :** (Color online) PFM images (6x6 $\mu m^2$ area) of the thinnest region of a $BiFeO_3$ polished single crystal.

**Figure 5.a :** Topography. Polishing scratches are visible.

**Figure 5.b :** Out-of-plane PFM phase image after writing "up" and "down" stripes (size: 4x0,5 μm). The out-of-plane polarization induced by negative voltage has the same direction as the spontaneous one.

**Figure 6 :** Local piezoelectric coefficient $d_{33}$ (not corrected by the tip effect) versus applied high voltage hysteresis at room temperature on a single crystal of $BiFeO_3$.

**Figure 7.a:** Charge current versus applied voltage of a 40 μm thick single crystal of $BiFeO_3$ at room temperature. The raw data are composed of a background leakage current superimposed to the relevant signal.

**Figure 7.b :** P-E hysteresis loop of the 40 μm thick single crystal of $BiFeO_3$ at room temperature.

**Figure 8 :** (Color online) Schematic antiferromagnetic structure of $BiFeO_3$ where the two AF sublattices are organized along a cycloidal spiral. The propagation vector **q** is along the direction [110] and the plane of spin rotation is (1-10).

**Figure 9 :** Magnetization curve versus applied magnetic field of the powder sample at room temperature.

**Figure 10 :** (Color online) Magnetization curves versus applied magnetic field of an unleached single crystal at several temperatures. The magnetic field is applied perpendicularly to the (012) plane.

**Figure 11 :** (Color Online) Magnetization curves versus temperature, under a field of 1 T, of an unleached (blue points) and of a leached (red points) single crystal of $BiFeO_3$. Lines are guides for the eye.

**Figure 12 :** Room temperature Mössbauer spectrum of a ceramic sample of $BiFeO_3$. The spectral asymmetry is clearly seen on the two outer lines. The line is a fit to an incommensurate spiral structure with a small anisotropy of the hyperfine interaction (see text).

**Table 1:** Four-circle X-ray diffraction data at 300K for a single crystal of $BiFeO_3$ using a $Mo_{K\alpha}$ radiation ($\lambda$ = 0.71073 Å).